\documentclass[letterpaper,twocolumn,10pt]{article}
\usepackage{usenix2021}
\usepackage{amsmath}
\usepackage{float}
\usepackage{graphicx}
\usepackage{xurl}

\begin{document}
\title{\Large \bf Revisiting Time, Clocks, and Synchronization}

\author{ {\rm Ying ZHANG}\\
 \rm \texttt{ying@smail.nju.edu.cn}}

\maketitle
\begin{abstract}
Sub-nanosecond precision clock synchronization over the packet network has been achieved by the White Rabbit protocol for a decade. However, few computer systems utilize such a technique. We try to attract more interest in the clock synchronization problem. We first introduce the basics of clock and synchronization in the time and frequency discipline. Then we revisit several related works, such as Google's Spanner, Huygens, FARMv2, DTP, and Sundial, explain why these works could be improved. Finally, we briefly discuss an independent time network approach towards low-cost and high-precision synchronization.
\end{abstract}

\section{Introduction}
Accurate time is everywhere in our daily life. 
The global navigation satellite systems (GNSS, namely GPS, Galileo, GLONASS, and Beidou) are, in fact, global high-precision clock dissemination systems. 
Navigation satellites are equipped with atomic clocks and are regularly synchronized with ground time laboratories. 
These ground laboratories can independently keep the accurate time. Meanwhile, they directly or indirectly synchronize with more than eighties of other time laboratories located globally.
All these time laboratories and the GNSS systems form the time service infrastructure for our society, by totally hundreds of atomic clocks and various time transfer channels, generating, keeping, and disseminating the Coordinated Universal Time (UTC) continuously and reliably for decades\cite{BIPMReport}.

Any location on the earth, such as datacenters in different cities, can use GNSS clocks to acquire the accurate time of fewer than 100 nanoseconds (ns) deviation from UTC\cite{NIST_GPS16}. 
However, computers inside a datacenter can only synchronize their clocks within several milliseconds when using the Network Time Protocol (NTP)\cite{NTPBook}.
The Precision Time Protocol (PTP, published by the IEEE 1588 standard) can synchronize computer clocks with microseconds precision. 
The latest version of PTP (IEEE 1588-2019\cite{PTPStd}) announced a high-precision profile, which is the standardization of the White Rabbit protocol \cite{WhiteRabbit} developed by the European Organization for Nuclear Research (CERN) around 2009, synchronizing at sub-nanosecond precision over 1 Gbps Ethernet with optical fibers.
\footnote{White rabbit's hardware design is open-sourced at \url{https://ohwr.org/project/white-rabbit}}

We would assert that high-precision clock synchronization over the packet network has been solved by the White Rabbit protocol for a decade.
This protocol was initially applied in scientific research facilities and is gaining adoption in financial systems\cite{WRforFinance}. However, its adoption is still limited\cite{PTP20}. 
Requiring dedicated hardware may be one reason.

Google's Spanner\cite{Spanner} may be the most well-known system that uses physical clocks. 
Recently, there is work to get the accurate time for each computer by distributing the GNSS signal\cite{RethinkingIT}. 
There are also some works claimed to achieve high-precision synchronization by the SVM machine learning algorithm (Huygens\cite{Huygens}), by the high-performance network (FARMv2\cite{FARMv2}), or by modifying the physical layer of the Ethernet (DTP\cite{DTP}, and Sundial\cite{Sundial}), even without any atomic clock or GNSS clock.

We will first briefly introduce the basic principles of time and frequency\cite{ch41Time,ch42Freq}. 
We are to point out that both the clock generating stable frequency and the time transfer channel introducing little variation are essential for high-precision clock synchronization. 
However, computers and network devices are only installed with ordinary quartz crystal oscillators, and the path delay over the packet network varies inherently. Both of them are not good for clock synchronization. 
Then we discuss why some related works, such as Google's Spanner, Huygens, FARMv2, DTP, and Sundial, can be improved. 
Finally, we discuss possible approaches towards low-cost, high-precision clock synchronization.

\section{Time}
Time is a concept about the ordering of events. In physics, time is a continuous variable to describe motion. However, the measured time is always \textbf{discrete}. The instrument we measuring time is the clock, mainly composed of an oscillator and a counter.

In his 1978 seminal article\cite{Lamport78}, Lamport proposed the logical time to order events in a distributed system, where clocks cannot be precisely synchronized.

The "Anomalous Behavior" section of Lamport's article discussed the limitation of logical time: it may order events opposite to physical time. Lamport pointed out an essential property of physical time:

{\itshape "One of the mysteries of the universe is that it is possible to construct a system of physical clocks which, running quite independently of one another, will satisfy the Strong Clock Condition. We can therefore use physical clocks to eliminate anomalous behavior."}

Interestingly, half of the references  (two out of four) of Lamport's article are about the relativity theory. The relativity theory changes our intuitive concept of absolute time, but it does not deny accurate time. The relativity theory prevents different reference frames from agreeing on "simultaneity," but it does not reverse the causal order. 
The time in different reference frames can be transformed to the observer's reference frame by the Lorentz transformation or the more sophisticated general relativity transformation. 

\section{Clock}

It is mentioned above that a clock is composed of an oscillator and a counter. 
The oscillator is a frequency source that generating a periodic pulse signal.
Stability is the essential characteristic of the oscillator. 
Atomic clocks employ cesium, hydrogen, or rubidium atoms as their oscillators, reaching the stability of $10^{-11}-10^{-13}$\cite{ch42Freq}.
The atomic clock is formally called "frequency standard," emphasizing the high-stable frequency signal as its primary output. 
The counter is relatively simple. However, it takes effort to transfer and align the counters within clocks. 
Although a atomic clock is sophisticated, it is simple to disseminate its high-stable frequency signal by a phase-locked loop (PLL) circuit.
\footnote{A PLL is a feedback circuit that locks a tunable oscillator to the input frequency. The PLL can amplify, clean, multiply, and dived the input frequency signal. \url{https://en.wikipedia.org/wiki/Phase-locked_loop}}

\begin{figure}[h]
  \centering
  \includegraphics[width=1.5in]{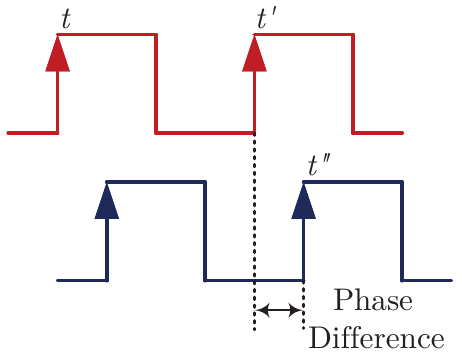}
  \caption{Timestamp and Phase Difference.}
\end{figure}

Computers use an integer as a timestamp to denote time. As shown in figure 1, a timestamp is captured at the rising edge of a pulse. The granularity of a timestamp is the period of the frequency signal. The fraction part of one period (phase difference) cannot be calculated by timestamps.

The output of atomic clocks and GNSS clocks has three parts: (i) a 10 MHz frequency signal; (ii) a time signal; and (iii) a time code signal. The rationale is that different type of signal carries a different characteristic of a clock. 

The 10 MHz frequency signal is, in fact, generated by an oven-controlled crystal oscillator (OCXO), which is disciplined to the atoms' or GNSS signal's frequency by a phase-locked loop with frequency multiplication. The frequency signal keeps the high-stability of clocks.

Dividing the 10 MHz frequency signal by $10^7$ times, we get a 1 Hz signal, which pulses once per second. 
A time signal, or one pulse per second (PPS) signal, is the 1 Hz signal with each pulse's rising edge accurately aligned to the beginning of UTC second.
The PPS signal represents the traceability accuracy to UTC.
Whenever a PPS signal is transferred to a downstream clock, the path delay must be compensated.

We need the time code signal to label each pulse of the PPS signal. The time code signal can be transmitted over a packet network, as there is almost an entire second duration to accomplish the transmission. 

It seems the one-second granularity of the PPS signal is too coarse. In fact, the time interval between a PPS signal and UTC is essential to the time signal, which is measured by a time interval counter (TIC), and usually has picoseconds resolution.
\footnote{The time-to-digital converter (TDC) is the core part of a TIC. 
\url{https://en.wikipedia.org/wiki/Time-to-digital_converter}.}

Clocks inside computers and network devices are driven by quartz crystal oscillators(XO for short).
This class of XOs costs less than one dollar for each.
They are not manufactured and calibrated for timekeeping and are not tunable. 
Usually, the actual output frequency of these XOs deviates from the nominal value in dozens of parts per million (ppm)
\footnote{That is, dozens of microseconds drift in one second.}, and they have no tuning port. 
Free-running XOs drift in the long-term due to environment temperature change, aging, and other factors. 
However, the short-term (in 1 second) stability of XOs is quite good.
When disciplined by a more stable clock using PLL, one XO can generate frequency almost as stable as the upstream clock.

The GNSS clock is a special kind of clock. 
In normal operation, a GNSS clock employs the GNSS signal to discipline its internal oscillator, making
the clock almost as stable as the GNSS system clocks in the long-term. 
The short-term stability of a GNSS clock is much determined by the quality of its internal oscillator.
A stable oscillator can filter out most GNSS signal noises.
More importantly, when the GNSS signal is blocked or spoofed,
\footnote{The antenna's position of a GNSS clock is fixed, so it is easy to detect signal blocking or spoofing by monitoring the calculated position.}
the internal oscillator is used to holdover.
Therefore, a high-end GNSS clock usually equips a rubidium atomic clock as its internal oscillator\cite{NIST_GPS08, TimeCard}.

\section{Synchronization} 

Synchronization (also called time transfer) determines the accuracy of a clock. That is, how far the clock could deviate from UTC.
To synchronize clocks, we must compensate for path delay. The key is not the absolute length of the delay but its variation.

Assuming we synchronize an atomic clock by the NTP protocol with a precision of one millisecond. It would exceed the variation of free-running in three years of the atomic clock.
\footnote{Suppose the stability of the atomic clock is $10^{-11}$\cite{ch42Freq}, then we get $10^{-3} / (10^{-11} \times 60 \times 60\times 24 \times 365) = 3.17$ years.}

We have one-way and two-way time transfer (synchronization) methods in principle\cite{ch41Time}. 

\subsection{One-way Time Transfer}
The one-way time transfer method is mainly GNSS-based, which is ideal in many aspects. It is a broadcast manner. The receivers calculate the path delay on their own, not need to interact with satellites. The receivers' amount is unlimited. Moreover, the synchronization precision is within 100 ns and can be improved with advanced receivers. 

The receiver can get the satellite $i$'s position $(x_i,y_i,z_i)$, as well as the sending time $t_i$ from its signal. The receiver measures the signal transmitting delay $\tau_i$ using its internal clock, which has an offset $\Delta$ to the GNSS time.
Let $(x,y,z)$ be the receiver's position, we get
\begin{equation}
  (\tau_i-\Delta)\cdot c=\sqrt{(x_i-x)^2+(y_i-y)^2+(z_i-z)^2}
\end{equation}
where $c$ is the light speed in the air.
The value of $(x,y,z)$ and $\Delta$ are unknown, which could be solved by four equations like this, receiving signals from four satellites. Then we can adjust the receiver's clock by $\Delta$.

Please note the GNSS antenna is connected to the receiver through a cable.
The signal speed in a cable is slower than in the air. Moreover, the signal does not travel along the cable in a straight line; we cannot calculate this part of the distance by the right-side formula in equation (1).
Thus, the cable's length from the antenna to the receiver must be measured and taken into account.

\subsection{Two-way Time Transfer} 
\begin{figure}[ht]
  \centering
  \includegraphics[width=3in]{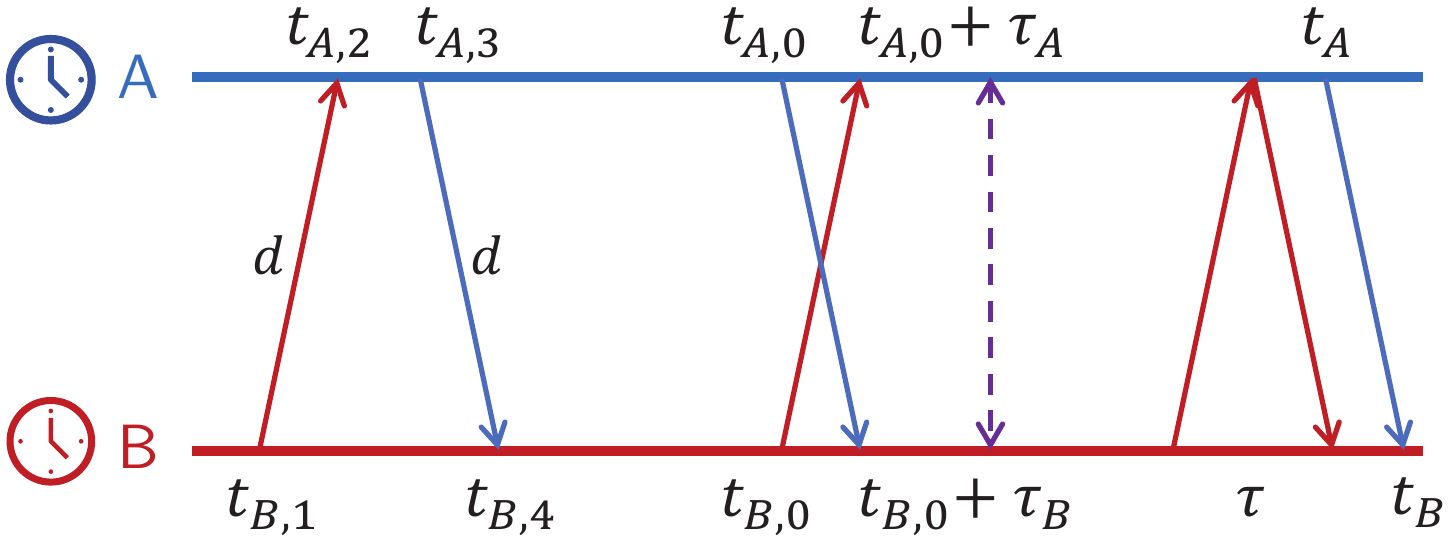}
  \caption{Variants of the Two-way Time Transfer Method.}
\end{figure}

The two-way time transfer method transmits the time signal between two clocks bidirectionally. 
This method assumes that the path is symmetric. That is, the forward path delay is equal to the backward path delay.
This is the method used in NTP, PTP, and the two-way satellite time transfer (TWSTT) method synchronizing time laboratories.
\footnote{The TWSTT method uses a geostationary satellite, not a GNSS satellite. In a short distance, a bidirectional optical fiber can be used instead of the satellite channel.}

Figure 2 shows three variants of the two-way time transfer method.
The left part in figure 2 is used by NTP and PTP. There is a request and a response, involving four timestamps.
The timestamps $t_{B,1}$ and $t_{B,4}$ are recorded by clock B, and the timestamps $t_{A,2}$ and $t_{A,3}$ are recorded by clock A. Let $\delta=t_B-t_A$ be the offset between clock A and B, and $d$ be the forward and backward path delay between A and B.
We have $t_{A,2}=t_{B,1}+d-\delta$ and $t_{B,4}=t_{A,3}+d+\delta$. Then we get 
$$d=[(t_{B,4}-t_{B,1})-(t_{A,3}-t_{A,2})]/2,$$
 and 
$$\delta=[(t_{B,4}-t_{A,3})+(t_{B,1}-t_{A,2})]/2.$$

The variant in the middle of figure 2 is the TWSTT method. Both clock A and B start to send a time signal to each other on the same time $t_{A,0}$ and $t_{B,0}$ according to their own clock (note that $t_{B,0}=t_{A,0}$), then measure the time intervals $\tau_A$ and $\tau_B$ on receiving signals by a time interval counter (also note that $t_{B,0}+\tau_{B}=t_{A,0}+d+\delta$, and $t_{A,0}+\tau_{A}=t_{B,0}+d-\delta$), 
exchange the measured intervals over a packet network, finally calculate 
 $$d=(\tau_B+\tau_A)/2 \text{, and } \delta=(\tau_B-\tau_A)/2.$$

The last variant is the round-trip (also called loop-back) method. The time signal sent by clock B is reflected instantly by clock A, the total interval $\tau$ is measured, and the path delay is 
$$d=\tau/2.$$ 
Whenever clock A sends a timestamp $t_{A}$ to clock B, the latter records its timestamp $t_{B}$ on receiving, and calculates the clock offset 
$$\delta=t_{B}-t_{A}-d.$$

\subsection{The White Rabbit Protocol}
The White Rabbit protocol \cite{WhiteRabbit, PTPStd} is an integration of PTPv2 (IEEE 1588-2008), Synchronous Ethernet (SyncE),\footnote{\url{https://en.wikipedia.org/wiki/Synchronous_Ethernet}.}
 time interval measurement, and Pre-calibration.
\begin{itemize}
  \item \textbf{PTPv2} requires switches to act as boundary clocks (BC) or transparent clocks (TC). A BC synchronizes its internal clock to the upstream clock, while a TC corrects timestamp packets with the internal queuing delay. PTPv2 records timestamps by hardware.
  \item \textbf{SyncE} synchronizes the quartz crystal oscillators' frequency (syntonization) inside network devices, making it possible to discipline these oscillators by an external high-stable clock. 
  \item \textbf{Time interval measurement} completes the fraction part (the phase difference in figure 1) of the timestamp. The carrier frequency signal is looped back to the sender through a single strand of optical fiber.\footnote{Usually, the Ethernet uses a pair of fibers for one link, one strand for each direction. \url{https://en.wikipedia.org/wiki/Ethernet_physical_layer}}
  The phase difference between the forward and backward carrier frequencies is measured at the sender. 
  \item \textbf{Pre-calibration}. The White Rabbit protocol uses light signals of different wavelengths for different directions through a single strand of optical fiber. The dispersion of light makes the forward and backward path through the same fiber asymmetric. Thus a calibration is needed to correct the asymmetry. After all, we cannot tell whether things work well without calibration.
\end{itemize}

The White Rabbit protocol works mostly at the physical layer, making it as deterministic as possible.

\section{Revisiting Related Works}

\textbf{Google's Spanner}\cite{Spanner}.
Someone would consider Google's Spanner to be the state of the art for clock synchronization at then\cite{CACMNow}. In fact, the PTPv2 protocol was published in 2008, achieved microseconds synchronization precision. The White Rabbit project was launched in 2009 and has been applied at CERN since around 2011, achieved sub-nanosecond precision. However, both PTP and White Rabbit were not discussed in the Spanner article.

\begin{itemize}
\item Spanner uses both GPS clocks and atomic clocks. It is mentioned that the cost of an atomic clock is of the same order as a GPS clock. This is because the high-end GPS clock equipped an atomic clock (usually a rubidium clock) internally for holdover, which is the most expensive part.
A standalone atomic clock is not enough. It needs to synchronize with the outside world regularly. However, the most convenient solution is the GNSS signal; alternative ones would be the TWSTT or optical fiber-based method as time laboratories do. It is rare to synchronize an atomic clock by the NTP or a software-based protocol.
\item A Spanner's time client polls from several time masters at different datacenters. Accessing time masters across datacenters contributes marginal to the reliability but significantly affects the precision when synchronizing over a long-distance network. 
\item It is unclear whether the error bound of synchronization $\epsilon$ has been measured with a reference clock, or it is just calculated by a software model. A model without physical verification is questionable.
\end{itemize}

\textbf{Huygens} \cite{Huygens} claimed a software-based clock synchronization system with 100 nanoseconds precision.
\begin{itemize}
  \item It is incorrect to interpret inequality (2) in the Huygens paper as the upper bound or lower bound of $\Delta_B-\Delta_A$. Denoted $\Delta_B-\Delta_A=\delta$ for short. According to the Huygens paper, the clocks in server A and B can be modeled as $C_A(t)=(1+\alpha)t+a$, $C_B(t)=(1+\beta)t+b$, where $t$ is the physical time, $\alpha$ and $\beta$ are the frequency bias of clock A and B, $a$ and $b$ are the time offset of clock A and B respectively. Let $d$ be the path delay from server A to B, and $d'$ the path delay from B to A. 
  
  The inequality (2) can be simplified to $\delta-d'<\delta<\delta+d$. Eliminating $\delta$, we get $-d'<0<d$, which is always held and has nothing to do with $\delta$. Actually, $\delta=\Delta_B-\Delta_A$ is approximately the offset between clock A and B. Nothing can bound the offset of two free-running clocks. The `forbidden zone' in figure 4 in the Huygens paper comes from the fact that few packet probes can travel faster than the shortest path delay.
  \item Huygens corrects the clock frequencies' discrepancy in the short-term by the SVM algorithm and exploiting a natural network effect. However, all the ordinary quartz crystal oscillators that driven computer clocks drift with temperature in the long-term. The drift cannot be limited unless these oscillators are synchronized with a more stable clock. 
\end{itemize}

\textbf{FARMv2}\cite{FARMv2} uses RDMA to exchange timestamps hundreds of thousand times per second, claiming it reached tens of microseconds synchronization precision without an atomic clock or a GNSS clock. It suffers the same long-term drift problem as Huygens. 

\textbf{DTP}\cite{DTP} and \textbf{Sundial}\cite{Sundial} modified the physical layer hardware to send timestamp packets more frequently than PTP. However, they cannot surpass PTP with SyncE enabled, which synchronizes the quartz crystal oscillators at the Ethernet carrier frequency, and can further disciplines them with a more stable clock.


There is work acquiring time by \textbf{distributing the GNSS signal} to each computer inside a datacenter \cite{RethinkingIT}. 
As we discussed in section 4.1, the cable length must be considered. It is unclear whether they use cables of certain fixed length options or measure the cable length on the field. Anyway, we do not think this approach achieves better precision than the White Rabbit protocol.

\section{Towards Low-cost High-precision Synchronization}


The White Rabbit protocol can achieve sub-nanosecond clock synchronization even across hundreds of kilometers\cite{CLONETS}. The problem is that it requires dedicated bidirectional optical fibers, which are not always available.
Unless the time transfer techniques used by time laboratories, such as GNSS common view and satellite/optical fiber-based two-way time transfer \cite{ch41Time}, are available, GNSS clocks are still the best way to inter-datacenter synchronization.

We are unsure whether the White Rabbit protocol would get more adoption since its standardization by IEEE 1588-2019. Indeed, it is expensive to update existing network devices to adopt the White Rabbit protocol. Besides, there is no 40 Gbps or faster Ethernet devices supporting the White Rabbit protocol yet. 

Is it worth achieving sub-nanosecond precision? We cannot answer as there are few applications using physical clocks yet. The Time Appliances Project (TAP) \cite{TAP_OCP} proposed a Datacenter PTP Profile based on the IEEE 1588 standard recently, which is aimed at few microseconds level precision for 40 Gbps and faster Ethernet in modern datacenters. 

It would be ideal if we could get high-precision synchronization at a low cost. A possible approach is deploying an \textbf{independent time network} \cite{CACMIllusion} with a simplified White Rabbit protocol.
Recall that the White Rabbit protocol works mostly at the physical layer. Instead of calculated by timestamps, the time interval can be measured more accurately by the time interval counter, triggered by the pulse's edge. 
We could reduce the cost by removing most of the packet network-related functions. Security and reliability are also improved, as a pure time network is almost administrating-free. Besides, independence allows the packet network and time network to evolve at their own pace.

Recently, the Sirius project \cite{Sirius} proposed a passive optical network. Besides its advanced attributes of the packet network, Sirius provides deterministic paths, a natural fit for high-precision synchronization. Sirius has already employed precise synchronization to schedule packets and reduce the link receiver's data recovery time. We expect that Sirius would push both the packet and time network forward.

\section{Conclusion}
High-precision clock synchronization is technically achievable. We regard the cost to be an obstacle hindering the adoption. We first introduced the basics of clock and synchronization. We recommend readers to Lombardi's informative chapters \cite{ch41Time,ch42Freq}. We revisited several related works, such as Google's Spanner, Huygens, FARMv2, DTP, and Sundial, discussed why they could be improved. We proposed an approach of deploying an independent time network.

This paper is only a vague start. We hope to attract more interest in the clock synchronization problem.

\bibliographystyle{plain}
\bibliography{ref}

\end{document}